\documentclass[twocolumn,showpacs,twoside,10pt,prl,superscriptaddress]{revtex4-1}
\usepackage{graphicx}
\usepackage{epsfig}
\usepackage{epsf}
\usepackage{amssymb}
\usepackage{amsmath}
\usepackage{amsthm}
\usepackage{multirow}

\begin{document}

\title{ Wavelet-based fast time-resolved magnetic sensing with electronic spins in diamond }

\author{Nanyang Xu}
\email{nyxu@ustc.edu.cn}
\affiliation{School of Electronic Science and Applied Physics, Hefei University of Technology, Hefei,  Anhui, 230009, China}

\author{Fengjian Jiang}
\affiliation{School of Information Engineering, Huangshan University, Huangshan, Anhui, 245041, China}

\author{Yu Tian}
\affiliation{School of Electronic Science and Applied Physics, Hefei University of Technology, Hefei,  Anhui, 230009, China}

\author{Jianfeng Ye}
\affiliation{School of Information Engineering, Huangshan University, Huangshan, Anhui, 245041, China}

\author{Fazhan Shi}
\affiliation{Hefei National Laboratory for Physics Sciences at Microscale and Department of Modern Physics, University of Science and Technology of China, Hefei, 230026, China.}

\author{Haijiang Lv}
\email{luhj9404@mail.ustc.edu.cn}
\affiliation{School of Information Engineering, Huangshan University, Huangshan, Anhui, 245041, China}

\author{Ya Wang}
\email{wangya837@gmail.com}
\affiliation{3. Physikalisches Institut, Research Center SCOPE, and MPI for Solid State Research,
University of Stuttgart, Pfaffenwaldring 57, 70569 Stuttgart, Germany}

\author{J\"org Wrachtrup}
\affiliation{3. Physikalisches Institut, Research Center SCOPE, and MPI for Solid State Research,
University of Stuttgart, Pfaffenwaldring 57, 70569 Stuttgart, Germany}

\author{Jiangfeng Du}
\affiliation{Hefei National Laboratory for Physics Sciences at Microscale and Department of Modern Physics, University of Science and Technology of China, Hefei, 230026, China.}

\begin{abstract}
Time-resolved magnetic sensing is of great importance from fundamental studies to applications in physical and biological sciences. Recently the nitrogen-vacancy (NV) defect center in diamond has been developed as a promising sensor of magnetic field under ambient conditions. However the methods to reconstruct time-resolved magnetic field with high sensitivity are not yet fully developed. Here, we propose and demonstrate a novel sensing method based on spin echo, and Haar wavelet transform. Our method is exponentially faster in reconstructing time-resolved magnetic field with comparable sensitivity over existing methods. Further, the wavelet's unique features enable our method to extract information from the whole signal with only part of the measuring sequences. We then explore this feature for a fast detection of simulated nerve impulses. These results will be useful to time-resolved magnetic sensing with quantum probes at nano-scales.

\end{abstract}
\pacs{76.70.r, 03.65.Ta, 07.55.Ge, 76.30.Mi}
\maketitle                            
Sensing of weak signals with high spatial resolution is of great importance in diverse areas ranging from fundamental physics and material science to biological sciences. NV center in diamond has recently emerged as one multifunctional sensor with high sensitivity and nano-scale spatial resolution under ambient conditions. It is applied for sensing magnetic fields\cite{nv_nano_mag_nature_Joerg,nv_nano_mag_nature_lukin,Taylor_magnetometer,high-dynran_magnetometer,hanson_magnetometer2011,degen_rf_magnetometer,wpf_vec_magnetometer,rugar_magnetometer2015}, electric fields\cite{electric_sensing}, temperature\cite{nv_thermometry,nv_thermal,nano_thermometer} and
magnetic resonance imaging\cite{Joerg_nano_nmr,single_esr,quimag,nmr_imaging,nmr_imaging_rugar2007}.

\begin{figure}[ht]
\centering
\includegraphics[width=1\columnwidth]{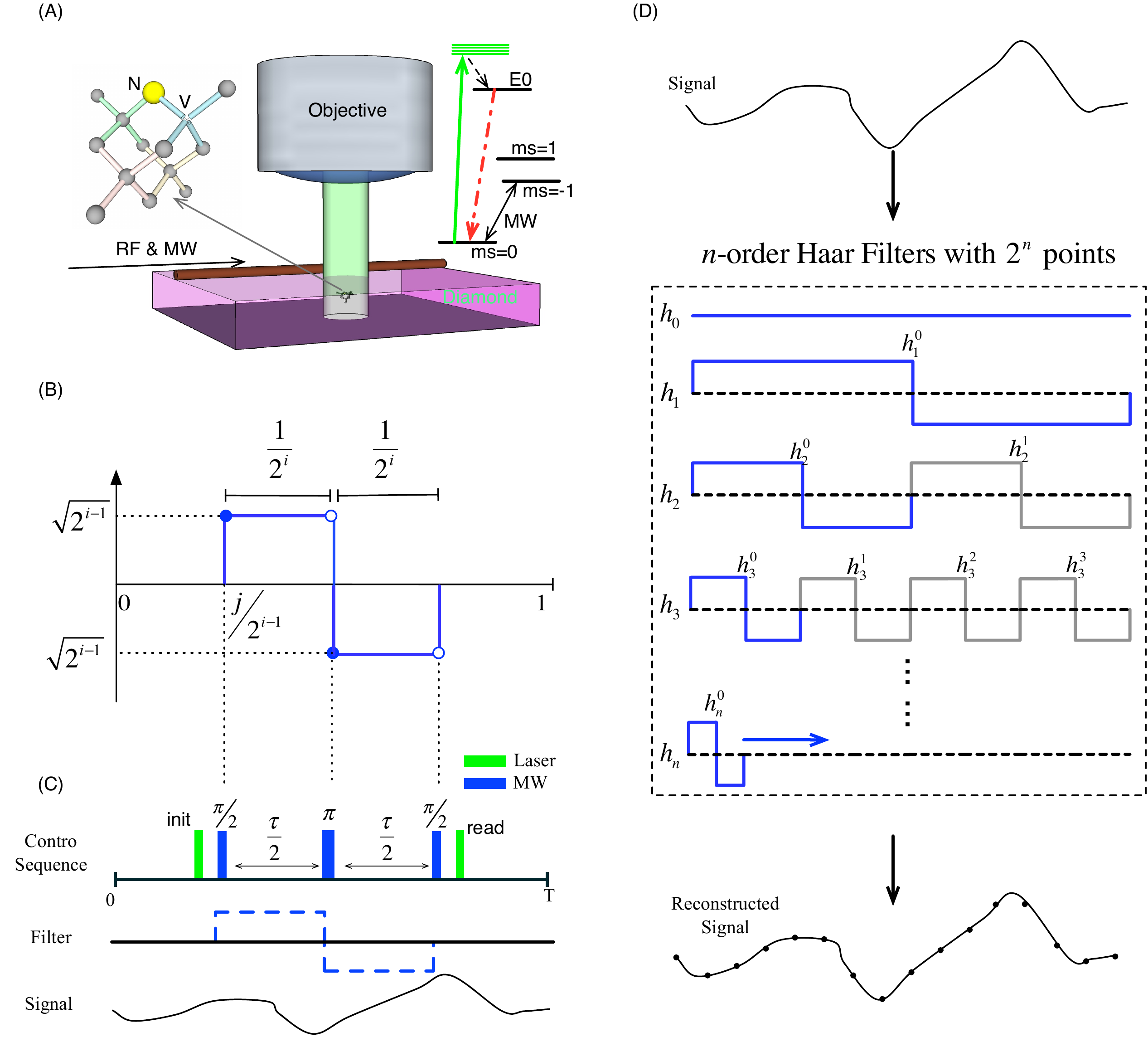}
\caption{(color online). Haar sensing method with NV center spin system. (A) The schematic of the setup. The NV center electron spin is initialized and readout via a confocal microscopy optical system, and coherently controlled by a resonate microwave to sense external temporal RF field. (B) General shape of Haar wavelet functions. (C) Spin echo sequence whose filter function matches well with Haar wavelet function. (D) Schematic of Haar sensing method. Temporal signal is measured by spin echo sequences associated with different Haar wavelet function to get the coefficients on corresponding Haar basis. Then an inverse Haar transform is performed with the coefficients to reconstruct the dynamics of the signal. A $2^{n}$-point reconstruction up to $n$-th order is demonstrated here.} 
\label{fig1}
\end{figure}

For NV based magnetic sensing applications, the basic idea is measuring magnetic field, for example, via Ramsey interferometry\cite{qmetro_review}: the electron spin is first prepared in an equal superposition of eigenstates $|0\rangle$ and $|1\rangle$ and obtains a relative phase in an external field $b(t)$. After an acquisition time $T$ the accumulated phase $\int_{0}^{T}{\gamma b(t)dt}$ results in a frequency shift that is measured optically where $\gamma$ is gyromagnetic ratio. This scheme is often called DC sensing scheme which could in principle extract the complete dynamics of the field required by general applications. By a successive monitoring of the resonance energy, either through increasing the acquisition period or sequential small acquisition steps, we can reconstruct the temporal magnetic field and extract its dynamics. However these reconstruction schemes are inefficient in sampling rate and also suffer from the electron spin's short coherence time (\emph{i.e.}, $T_2^{*}$) which limits the measurement sensitivity. Dynamical decoupling sequences\cite{du_nature_2009,dd_nv, opt_dd_nv, dd_nv_Joerg,dd_nv_hanson}, acting as a controllable frequency band-pass filter, could be used to enhance the coherence time by filtering out a large part of the noise\cite{noise_spectrum,noise_spectrum_suter,qu_lia}. High sensitivity of constant oscillating magnetic fields is achieved based on this method. But the temporal field's reconstruction is not straightforward.

To deal with this problem, recently Cooper et.al designed a group of decoupling sequences associated with the Wash functions\cite{walsh_sensing}. While protecting against dephasing noise, each Walsh decoupling sequence is measuring a Wash transform coefficient of temporal magnetic field on the corresponding Walsh basis. With the coefficients in different Walsh basis, an inverse Walsh transform finally reconstructs the temporal signal. For a $2^n$-point temporal signal reconstruction, $2^n$ coefficients need to be obtained through $2^n$ different decoupling sequences. This implies that an $O(2^n)$ runs of signal are required by Walsh method, which becomes resource consuming and slow reconstruction method in high orders (large n). Meanwhile, the pulse errors induced contrast loss and artifacts in decoupling sequence containing many pulses also limit the application of Walsh method in practice. A fast and easily implementable reconstruction method is therefore
more favorable for general applications. 

In this letter, we propose and experimentally demonstrate a novel method that can reconstruct a temporal signal exponentially faster with sensitivity comparable to the Walsh method. Specifically a $2^n$-point temporal signal is reconstructed with only $O(n)$ runs of signal with this new method. In particular, our reconstruction scheme is based on the Haar wavelet and is easily implemented with the standard spin echo technique in experiment. With the spin echo technique, our scheme also achieves a significant improvement in coherence time and sensitivity over Ramsey interferometry. Moreover our methods is able to extract information from the whole signal, which leads to even further less resource requirement. By exploring this unique feature, we finally apply this method into bio-sensing area and demonstrate a fast detection of nerve impulses.

The Haar wavelet \cite{haar_appl,haar_status} we are using here is the simplest form of wavelet functions\cite{wavelet_textbook}. Its mother function is a step function taking values 1 and -1, on [0,1/2) and [1/2,1) respectively, and 0 otherwise. Dilation and translation of this mother function define an orthonormal and complete basis $\{h_i^j$\}, where the $j$-th function $h_i^j$ of order $i$ for 
$j=\{0,\cdots,2^{(i-1)}-1\}$ is:
 $$
 h_i^j(x)=\left\{
 \begin{array}{lc}  
  2^{(i-1)/2},  & j/2^{i-1}\leqslant x< (j+\frac{1}{2})/2^{i-1} ;  \\  
  -2^{(i-1)/2},& (j+\frac{1}{2})/2^{i-1}\leqslant x< (j+1)/2^{i-1} ; \\  
 0, &  \text{other};
 \end{array}\right.
 $$ 
 and $h_0 \equiv 0$. 

Any function $f(x)$ defined over interval [0,1] with $f^2(x)$ integrable in the Lebesque sense could be decomposed in the Haar basis:
$$
f(x) = c_0 + \sum_{i=1}^{\infty}\sum_{j=0}^{2^{i-1}-1}{c_i^jh_i^j(x)},
$$
with coefficient
$$
c_i^j = \int_0^1{f(x)h_i^j(x)dx}.
$$

This means that with $2^n$ Haar coefficients a temporal field can be approximated up to the $i=n$-th order by:
$$
S_n(x) = c_0 + \sum_{i=1}^{n}\sum_{j=0}^{2^{i-1}-1}{c_i^jh_i^j(x)}.
$$

One property of the Haar wavelet is that all the basis have the same shape (see Fig.\ref{fig1}B) inherited from
the mother function. Such a shape perfectly matches the filter of spin echo sequence as shown in Fig.\ref{fig1}C, in which
the $\pi$ pulse effectively switches the sign of the temporal field from +1 to -1. The coefficient $c_i^j$ is therefore obtained by performing the spin echo sequence corresponding to the Haar wavelet $h_i^j$, through the relationship $c_i^j = 2^{-(i+1)/2}\pi\phi_i^j/(\tau_i\gamma)$, where $\phi_i^j$ is the phase acquired by the spin sensor in the spin echo sequence and $\tau_i$ is the measuring time of $i$-th order Haar sequence. Another important property of the Haar wavelet is its short-wavelike functions allowing all the $i$-th order coefficients $c_i^{j=0,...,2^i-1}$ to be successively measured, in principle, in one run of the temporal signal. Even taking into account the fast initialization and readout in each spin echo sequence, $O(n)$ runs of signal is enough to reconstruct a $2^n$-point temporal signal.

\begin{figure}[tbp]
\centering
\includegraphics[width=1\columnwidth]{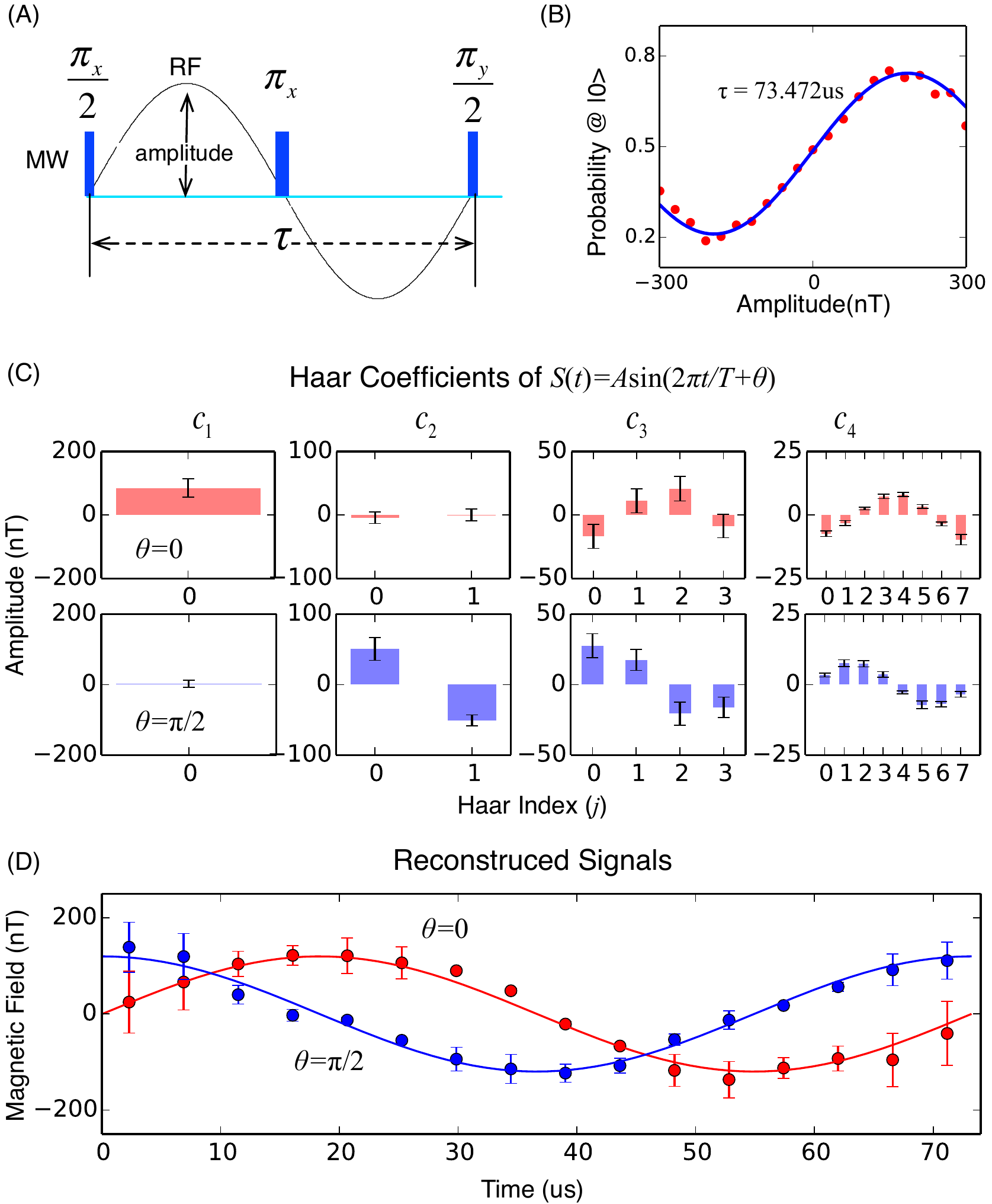}
\caption{(color online). Haar reconstruction of sinusoid signals. (A) Sequence for measuring RF fields for references, a spin echo sequence with an in-phase sinusoid RF field. Note that the two $\pi/2$ pulses are applied on different axes. (B) Relation between RF field's amplitude and NV electron spin's optical contrast. By varying the amplitude of RF field, the contrast change sinusoidally. The solid line is a fit of experiment points and used as a reference in Haar method.  (C) Haar coefficients get from the experiments up to fourth orders ($n=4$).  Error bars correspond to 95\% confidence intervals on the coefficients associated with the measured and fit errors. (D)16-point reconstruction of the signal, where the solid lines are theoretical expectations. Error bars correspond to the amplitude uncertainty of the reconstructed field, which is coming from the coefficients uncertainty shown in (C).} \label{fig2}
\end{figure}

We experimentally demonstrated this reconstruction method with a single NV center doped in a bulk diamond with natural $^{13}C$ abundance. The schematics of the setup and energy spectrum is shown in Fig.\ref{fig1}A. The negatively charged defect exhibits a ground state electronic spin triplet with about 2.87 GHz zero-field energy splitting between the $m_s = 0$ and $m_s = \pm1$
sub-levels. A static magnetic field is applied along the NV axis to remove the energy degeneracy between the $m_s = +1$ and
$m_s = -1$ sub-levels. We use $m_s=0$ and $m_s=-1$ sub-levels as a qubit sensor in the experiment.
A 532-nm green laser is used for the initialization and readout of the electron spin.
Coherent control of the spin is realized through the resonant microwave pulse radiated from a copper line mounted on the diamond. The temporal magnetic field is generated via an arbitrary wave generator (RF-AWG) and radiated from the same copper line. Details of the experimental parameters is in the supplementary information (SI).

\begin{figure}[tbp]
\centering
\includegraphics[width=1\columnwidth]{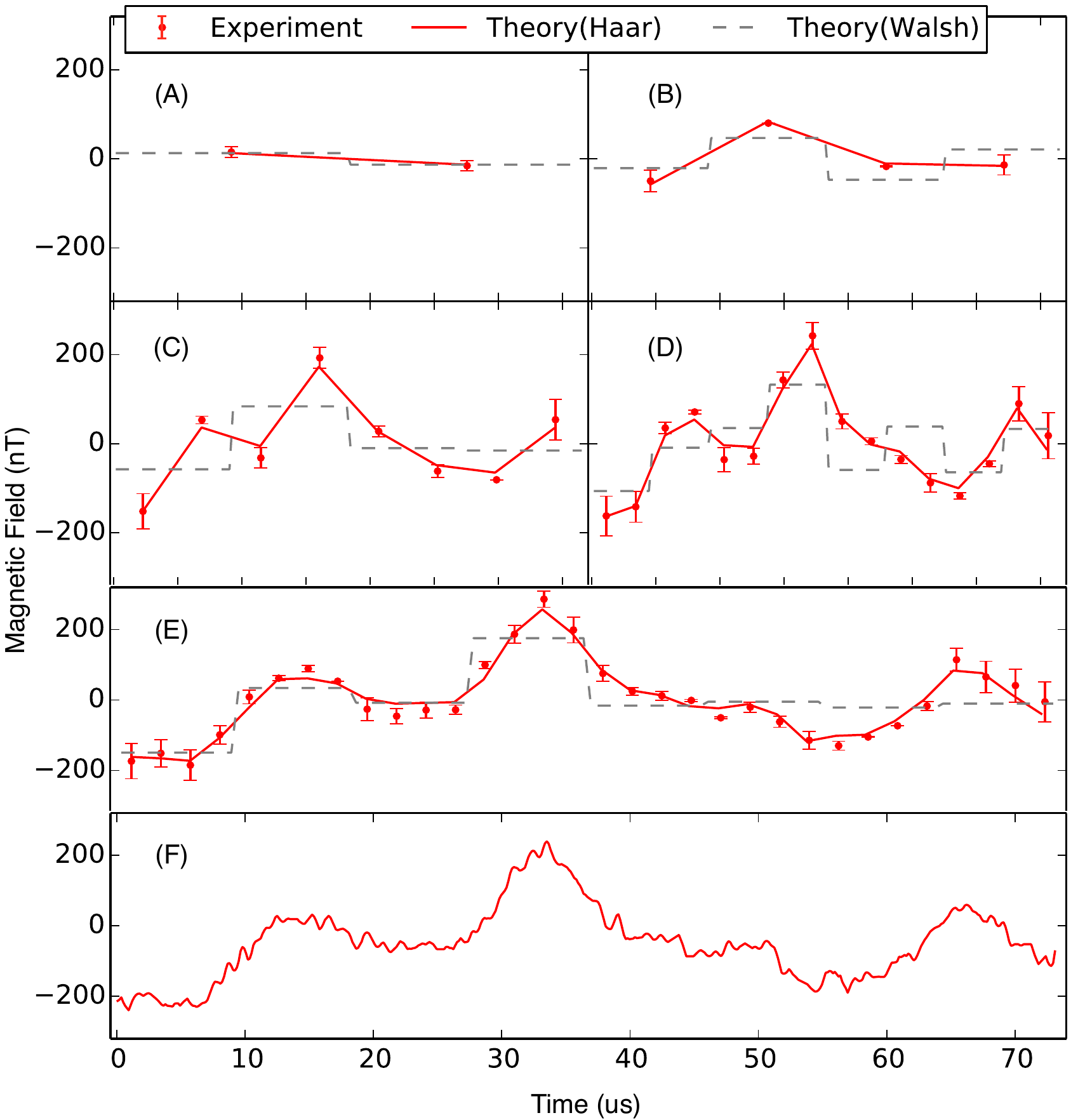}
\caption{(color online). Haar reconstruction of an arbitrary signal. The signal is reconstructed by measuring the Haar wavelet up to fifth order. (A)--(E) Reconstructed signal with up to order $n$ as $n$ changes from 1 to 5. Lines are theoretical expectations. The signal views from vague to clear as $n$ increases, which shows a progressive property of Haar method. Error bars correspond to the amplitude uncertainty of the reconstructed field obtained by propagation of the errors from experimental results and fitting of the parameters.  (F) The original signal output via RF-AWG. All the diagrams showing coefficients and reconstructed fields are manipulated by matplotlib\cite{matplotlib} in this letter. } \label{fig3}
\end{figure}

\begin{figure}[tbp]
\centering
\includegraphics[width=1\columnwidth]{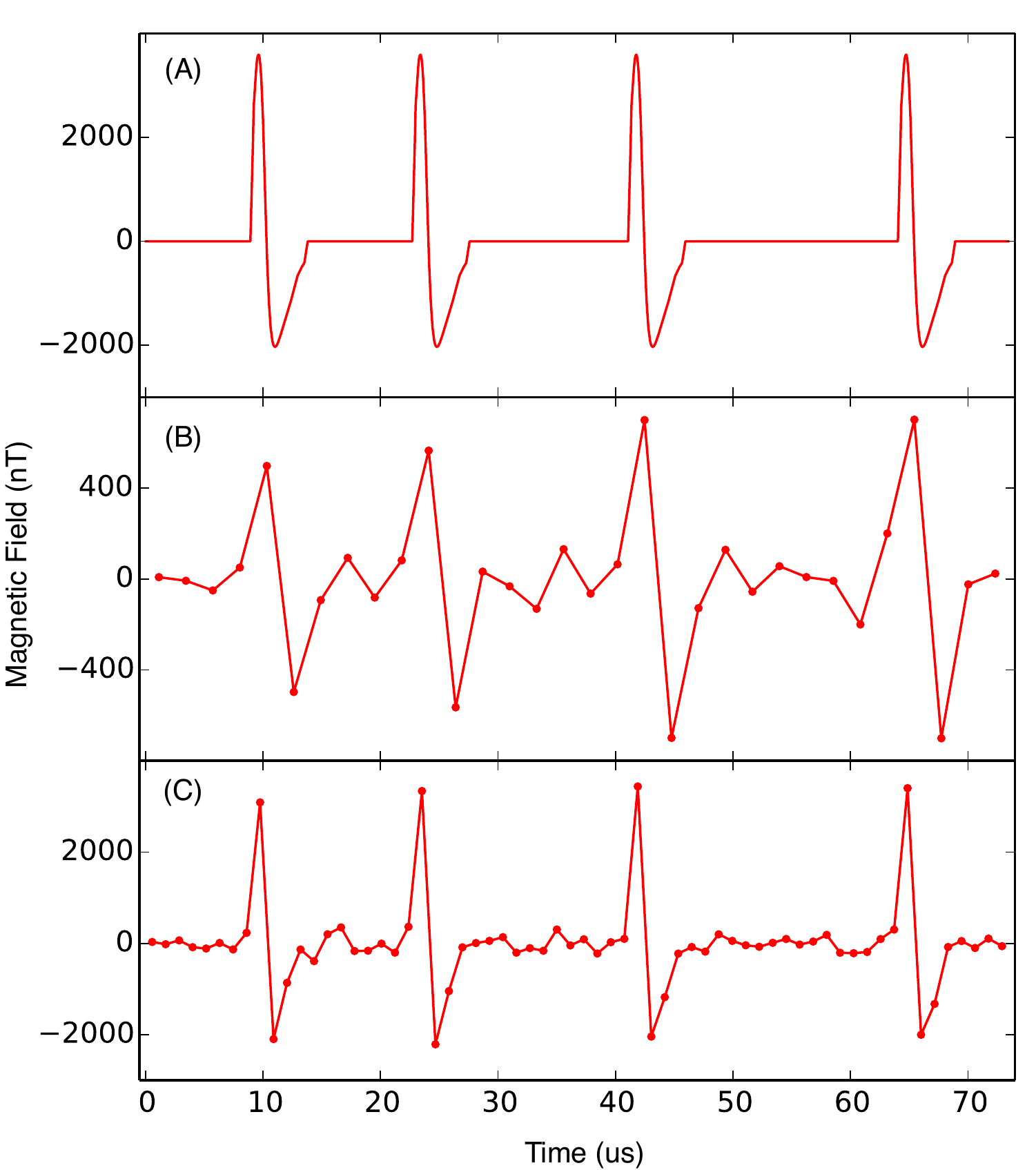}
\caption{(color online). Haar detection of nerve impulses. A) Simulated signal from a neuron. Each impulse is obtained from deviation of the action potential calculated by NEURON numerical simulation environment \cite{neuron_sim}. Four nerve impulse are arranged randomly as a train to be measured. B) Haar detection of the signal with fifth order Haar sequence. C) Haar detection of the signal with fifth and sixth order Haar sequences. The position and phase of the impulses are clearly represented by Haar method. Each sequence was looped $10^6$ times and the total acquisition time for measuring all the Haar coefficients was about 15 minutes.} \label{fig4}
\end{figure}

Prior to Haar sensing experiments, we need to characterize the effective RF magnetic field with spin echo experiment as references for each order. By applying an in-phase sinusoidal RF field and varying the amplitude\cite{ nv_nano_mag_nature_Joerg, nv_nano_mag_nature_lukin}, the obtained phase on the electron spin generates a sinusoid modulation on its optical contrast as in Fig.\ref{fig2}B. The fit line of the points is used as a reference for each order and more results are in SI.

We first reconstruct a monochromatic sinusoid field $S(t) = A\sin(2\pi t/T + \theta)$ by measuring its Haar coefficients up to fourth order ($n=4$). Respectively we chose two quadratic waveforms with $\theta = 0 $ and $\pi/2$. The measured Haar coefficients are shown in Fig.\ref{fig2}C and are arranged in columns with $i$. The reconstructed 16-point waveform by an inverse Haar transform matches well with theoretical expectations as is shown in Fig.\ref{fig2}D. This also shows that Haar sensing method is phase sensitive unlike original AC sensing methods, as it can reconstruct quadratic signals with the same sequences.

We then reconstruct a more complicated waveform, chosen randomly from a photograph (details in SI). The signal is measured up to fifth order (n=5). By an increasing reconstruction with n from 1 to 5, the reconstructed field is shown in Fig.\ref{fig3} which matches well with theoretical expectation. This implies the Haar method works universally and precisely for arbitrary waveforms in a general way. Furthermore, we compare the performance with Walsh method plotted as grey dashed lines in Fig.\ref{fig3} for the same order of experiments, showing that Haar method is much faster than Walsh method to reconstruct the same signal. Note that, Fig.\ref{fig3} shows an important feature of wavelet function, that the sensing process is progressive from vague to clear as the sensing order increases, which may be quite useful in quantum imaging areas\cite{nv_nano_mag_nature_Joerg, Taylor_magnetometer,qimag_degen,qimag_domainwall,lukin_imaging,yacoby_imaging,lukin_farfield-imaging,joerg_imaging2013,high_dynrange_imaging,nv_imag_projreconstr}.

In addition to a universal information extraction method, wavelet transform is widely used in modern data processing like imaging processing, data coding and pattern recognition\cite{haar_appl, haar_status}.
One main and unique feature is that wavelet transform could extract interesting information with only part of coefficients, which
makes wavelet transform different from other information extraction methods like Fourier transform and Walsh transform. 
Our wavelet reconstruction method also inherits this feature which is great helpful in fast information extraction, e.g., in sensing biological signals\cite{nv_fluctmag,nv_ionch,nv_neuron,nv_imag_ensem}. To show this point, we simulate a typical nerve impulses signal obtained from NEURON environment\cite{neuron_sim}. As is shown Fig.\ref{fig4}A, the test signal is a random train of the nerve impulses. To detect the signal, we apply a part of Haar sequences that have comparable width with the impulse. Fig.\ref{fig4}B and C show the experimental results using only one order ($n$=5) and two orders ($n$= 5 and 6) with Haar methods. The nerve impulses are clearly detected, which implies that, our reconstruction method could detect interested information faster and with even less runs of signals.

Finally we would like to discuss the sensitivity of Haar method. The amplitude resolution of the Haar method,
$(\delta b_N)_{haar} = \sqrt{\sum_{i=1}^{N}\delta b^2_i}$,  gives the smallest variation
of the reconstructed field that can be measured from order n ($N=2^n$). For each spin echo sequence used in Haar method, 
the minimal detectable magnetic field from M measurements is $\delta b_i^j \sim 1/{\sqrt{M\tau_i T_2}}$\cite{nv_nano_mag_nature_lukin}, where $\tau_i=T/2^{i-1}$ is the measuring time in $i$-th order experiment. A straightforward calculation would yield the amplitude resolution of Haar method, $(\delta b_N)_{haar}\sim\sqrt{\frac{N^2-1}{3M T T_2}}$,  
which provides a gain in sensitivity of $\sqrt{{T_2}/{T_2^*}}\cdot \sqrt{{3}/{n}}$ over sequential Ramsey interferometry methods 
that perform N successive amplitude measurements over the intervals of length $\tau = T/N\sim T_2^*$.
For Walsh method the enhancement is $\sqrt{{T_2}/{T_2^*}}$, which becomes exactly the number $\sqrt{N}$ given in \cite{walsh_sensing} if $T_2 \sim T$.
Hence, up to a $N=2^{10}$-point reconstruction (tenth order), the sensitivity of Haar method is comparable to the Walsh method (details in SI). 

To be concluded, we perform a proof-of-principle demonstration of a novel reconstruction method for quantum sensing of temporal signal based on Haar wavelet transform. Our method is shown to reconstruct temporal signals with exponentially less runs of signals.
and could be even less if only interested information are extracted. Since the demonstration is realized in single electron spin in NV center, an additional $10^6$ runs of signals are required for individual measurement accumulation in current experiment. But this requirement can be easily removed by performing the same measurements with an NV ensemble which could provide over $10^{11}$ spins in $10^{-4}$ mm$^3$ volume\cite{nv_ensemble_magnetometer,budker_cavity}. The standard spin echo technique used in our method not only makes it more robust and easily realized in practical application, but also readily applicable in other quantum sensing systems like trapped ions, semiconductors or quantum dots.

This work was supported by 973 Program (2013CB921800), NNSFC (61376128, 11227901, 11305074, 11135002, 11275083), DFG ( For 1493, SPP 1601, SFB/TR21), EU (SIQS, SQUTEC, DIADEMS) and CAS (XDB01030400).

%\bibliography{papers}

\begin{thebibliography}{46}%
\makeatletter
\providecommand \@ifxundefined [1]{%
 \@ifx{#1\undefined}
}%
\providecommand \@ifnum [1]{%
 \ifnum #1\expandafter \@firstoftwo
 \else \expandafter \@secondoftwo
 \fi
}%
\providecommand \@ifx [1]{%
 \ifx #1\expandafter \@firstoftwo
 \else \expandafter \@secondoftwo
 \fi
}%
\providecommand \natexlab [1]{#1}%
\providecommand \enquote  [1]{``#1''}%
\providecommand \bibnamefont  [1]{#1}%
\providecommand \bibfnamefont [1]{#1}%
\providecommand \citenamefont [1]{#1}%
\providecommand \href@noop [0]{\@secondoftwo}%
\providecommand \href [0]{\begingroup \@sanitize@url \@href}%
\providecommand \@href[1]{\@@startlink{#1}\@@href}%
\providecommand \@@href[1]{\endgroup#1\@@endlink}%
\providecommand \@sanitize@url [0]{\catcode `\\12\catcode `\$12\catcode
  `\&12\catcode `\#12\catcode `\^12\catcode `\_12\catcode `\%12\relax}%
\providecommand \@@startlink[1]{}%
\providecommand \@@endlink[0]{}%
\providecommand \url  [0]{\begingroup\@sanitize@url \@url }%
\providecommand \@url [1]{\endgroup\@href {#1}{\urlprefix }}%
\providecommand \urlprefix  [0]{URL }%
\providecommand \Eprint [0]{\href }%
\providecommand \doibase [0]{http://dx.doi.org/}%
\providecommand \selectlanguage [0]{\@gobble}%
\providecommand \bibinfo  [0]{\@secondoftwo}%
\providecommand \bibfield  [0]{\@secondoftwo}%
\providecommand \translation [1]{[#1]}%
\providecommand \BibitemOpen [0]{}%
\providecommand \bibitemStop [0]{}%
\providecommand \bibitemNoStop [0]{.\EOS\space}%
\providecommand \EOS [0]{\spacefactor3000\relax}%
\providecommand \BibitemShut  [1]{\csname bibitem#1\endcsname}%
\let\auto@bib@innerbib\@empty
%</preamble>
\bibitem [{\citenamefont {Balasubramanian}\ \emph {et~al.}(2008)\citenamefont
  {Balasubramanian}, \citenamefont {Chan}, \citenamefont {Kolesov} \emph
  {et~al.}}]{nv_nano_mag_nature_Joerg}%
  \BibitemOpen
  \bibfield  {author} {\bibinfo {author} {\bibfnamefont {G.}~\bibnamefont
  {Balasubramanian}}, \bibinfo {author} {\bibfnamefont {I.~Y.}\ \bibnamefont
  {Chan}}, \bibinfo {author} {\bibfnamefont {R.}~\bibnamefont {Kolesov}},
  \emph {et~al.},\ }\href@noop {} {\bibfield  {journal} {\bibinfo  {journal}
  {Nature}\ }\textbf {\bibinfo {volume} {455}},\ \bibinfo {pages} {648}
  (\bibinfo {year} {2008})}\BibitemShut {NoStop}%
\bibitem [{\citenamefont {Maze}\ \emph {et~al.}(2008)\citenamefont {Maze},
  \citenamefont {Stanwix}, \citenamefont {Hodges} \emph
  {et~al.}}]{nv_nano_mag_nature_lukin}%
  \BibitemOpen
  \bibfield  {author} {\bibinfo {author} {\bibfnamefont {J.~R.}\ \bibnamefont
  {Maze}}, \bibinfo {author} {\bibfnamefont {P.~L.}\ \bibnamefont {Stanwix}},
  \bibinfo {author} {\bibfnamefont {J.~S.}\ \bibnamefont {Hodges}},  \emph
  {et~al.},\ }\href@noop {} {\bibfield  {journal} {\bibinfo  {journal}
  {Nature}\ }\textbf {\bibinfo {volume} {455}},\ \bibinfo {pages} {644}
  (\bibinfo {year} {2008})}\BibitemShut {NoStop}%
\bibitem [{\citenamefont {Taylor}\ \emph {et~al.}(2008)\citenamefont {Taylor},
  \citenamefont {Cappellaro}, \citenamefont {Childress} \emph
  {et~al.}}]{Taylor_magnetometer}%
  \BibitemOpen
  \bibfield  {author} {\bibinfo {author} {\bibfnamefont {J.~M.}\ \bibnamefont
  {Taylor}}, \bibinfo {author} {\bibfnamefont {P.}~\bibnamefont {Cappellaro}},
  \bibinfo {author} {\bibfnamefont {L.}~\bibnamefont {Childress}},  \emph
  {et~al.},\ }\href@noop {} {\bibfield  {journal} {\bibinfo  {journal} {Nat.
  Phys.}\ }\textbf {\bibinfo {volume} {4}},\ \bibinfo {pages} {810} (\bibinfo
  {year} {2008})}\BibitemShut {NoStop}%
\bibitem [{\citenamefont {Waldherr}\ \emph {et~al.}(2012)\citenamefont
  {Waldherr}, \citenamefont {Beck}, \citenamefont {Neumann} \emph
  {et~al.}}]{high-dynran_magnetometer}%
  \BibitemOpen
  \bibfield  {author} {\bibinfo {author} {\bibfnamefont {G.}~\bibnamefont
  {Waldherr}}, \bibinfo {author} {\bibfnamefont {J.}~\bibnamefont {Beck}},
  \bibinfo {author} {\bibfnamefont {P.}~\bibnamefont {Neumann}},  \emph
  {et~al.},\ }\href@noop {} {\bibfield  {journal} {\bibinfo  {journal} {Nat
  Nanotechnol}\ }\textbf {\bibinfo {volume} {7}},\ \bibinfo {pages} {105}
  (\bibinfo {year} {2012})}\BibitemShut {NoStop}%
\bibitem [{\citenamefont {de~Lange}\ \emph {et~al.}(2011)\citenamefont
  {de~Lange}, \citenamefont {Riste}, \citenamefont {Dobrovitski},\ and\
  \citenamefont {Hanson}}]{hanson_magnetometer2011}%
  \BibitemOpen
  \bibfield  {author} {\bibinfo {author} {\bibfnamefont {G.}~\bibnamefont
  {de~Lange}}, \bibinfo {author} {\bibfnamefont {D.}~\bibnamefont {Riste}},
  \bibinfo {author} {\bibfnamefont {V.~V.}\ \bibnamefont {Dobrovitski}}, \ and\
  \bibinfo {author} {\bibfnamefont {R.}~\bibnamefont {Hanson}},\ }\href@noop {}
  {\bibfield  {journal} {\bibinfo  {journal} {Phys Rev Lett}\ }\textbf
  {\bibinfo {volume} {106}} (\bibinfo {year} {2011})}\BibitemShut {NoStop}%
\bibitem [{\citenamefont {Loretz}\ \emph {et~al.}(2013)\citenamefont {Loretz},
  \citenamefont {Rosskopf},\ and\ \citenamefont
  {Degen}}]{degen_rf_magnetometer}%
  \BibitemOpen
  \bibfield  {author} {\bibinfo {author} {\bibfnamefont {M.}~\bibnamefont
  {Loretz}}, \bibinfo {author} {\bibfnamefont {T.}~\bibnamefont {Rosskopf}}, \
  and\ \bibinfo {author} {\bibfnamefont {C.~L.}\ \bibnamefont {Degen}},\
  }\href@noop {} {\bibfield  {journal} {\bibinfo  {journal} {Phys Rev Lett}\
  }\textbf {\bibinfo {volume} {110}} (\bibinfo {year} {2013})}\BibitemShut
  {NoStop}%
\bibitem [{\citenamefont {Wang}\ \emph {et~al.}(2015)\citenamefont {Wang},
  \citenamefont {Yuan}, \citenamefont {Huang} \emph
  {et~al.}}]{wpf_vec_magnetometer}%
  \BibitemOpen
  \bibfield  {author} {\bibinfo {author} {\bibfnamefont {P.}~\bibnamefont
  {Wang}}, \bibinfo {author} {\bibfnamefont {Z.}~\bibnamefont {Yuan}}, \bibinfo
  {author} {\bibfnamefont {P.}~\bibnamefont {Huang}},  \emph {et~al.},\
  }\href@noop {} {\bibfield  {journal} {\bibinfo  {journal} {Nat Commun}\
  }\textbf {\bibinfo {volume} {6}} (\bibinfo {year} {2015})}\BibitemShut
  {NoStop}%
\bibitem [{\citenamefont {Mamin}\ \emph {et~al.}(2014)\citenamefont {Mamin},
  \citenamefont {Sherwood}, \citenamefont {Kim} \emph
  {et~al.}}]{rugar_magnetometer2015}%
  \BibitemOpen
  \bibfield  {author} {\bibinfo {author} {\bibfnamefont {H.~J.}\ \bibnamefont
  {Mamin}}, \bibinfo {author} {\bibfnamefont {M.~H.}\ \bibnamefont {Sherwood}},
  \bibinfo {author} {\bibfnamefont {M.}~\bibnamefont {Kim}},  \emph {et~al.},\
  }\href@noop {} {\bibfield  {journal} {\bibinfo  {journal} {Phys Rev Lett}\
  }\textbf {\bibinfo {volume} {113}} (\bibinfo {year} {2014})}\BibitemShut
  {NoStop}%
\bibitem [{\citenamefont {Dolde}\ \emph {et~al.}(2011)\citenamefont {Dolde},
  \citenamefont {Fedder}, \citenamefont {Doherty} \emph
  {et~al.}}]{electric_sensing}%
  \BibitemOpen
  \bibfield  {author} {\bibinfo {author} {\bibfnamefont {F.}~\bibnamefont
  {Dolde}}, \bibinfo {author} {\bibfnamefont {H.}~\bibnamefont {Fedder}},
  \bibinfo {author} {\bibfnamefont {M.~W.}\ \bibnamefont {Doherty}},  \emph
  {et~al.},\ }\href@noop {} {\bibfield  {journal} {\bibinfo  {journal} {Nat.
  Phys.}\ }\textbf {\bibinfo {volume} {7}},\ \bibinfo {pages} {459} (\bibinfo
  {year} {2011})}\BibitemShut {NoStop}%
\bibitem [{\citenamefont {Toyli}\ \emph {et~al.}(2013)\citenamefont {Toyli},
  \citenamefont {Charles}, \citenamefont {Christle} \emph
  {et~al.}}]{nv_thermometry}%
  \BibitemOpen
  \bibfield  {author} {\bibinfo {author} {\bibfnamefont {D.~M.}\ \bibnamefont
  {Toyli}}, \bibinfo {author} {\bibfnamefont {F.}~\bibnamefont {Charles}},
  \bibinfo {author} {\bibfnamefont {D.~J.}\ \bibnamefont {Christle}},  \emph
  {et~al.},\ }\href@noop {} {\bibfield  {journal} {\bibinfo  {journal}
  {Proceedings of the National Academy of Sciences}\ }\textbf {\bibinfo
  {volume} {110}},\ \bibinfo {pages} {8417} (\bibinfo {year}
  {2013})}\BibitemShut {NoStop}%
\bibitem [{\citenamefont {Kucsko}\ \emph {et~al.}(2013)\citenamefont {Kucsko},
  \citenamefont {Maurer}, \citenamefont {Yao} \emph {et~al.}}]{nv_thermal}%
  \BibitemOpen
  \bibfield  {author} {\bibinfo {author} {\bibfnamefont {G.}~\bibnamefont
  {Kucsko}}, \bibinfo {author} {\bibfnamefont {P.~C.}\ \bibnamefont {Maurer}},
  \bibinfo {author} {\bibfnamefont {N.~Y.}\ \bibnamefont {Yao}},  \emph
  {et~al.},\ }\href@noop {} {\bibfield  {journal} {\bibinfo  {journal}
  {Nature}\ }\textbf {\bibinfo {volume} {500}},\ \bibinfo {pages} {54}
  (\bibinfo {year} {2013})}\BibitemShut {NoStop}%
\bibitem [{\citenamefont {Neumann}\ \emph {et~al.}(2013)\citenamefont
  {Neumann}, \citenamefont {Jakobi}, \citenamefont {Dolde} \emph
  {et~al.}}]{nano_thermometer}%
  \BibitemOpen
  \bibfield  {author} {\bibinfo {author} {\bibfnamefont {P.}~\bibnamefont
  {Neumann}}, \bibinfo {author} {\bibfnamefont {I.}~\bibnamefont {Jakobi}},
  \bibinfo {author} {\bibfnamefont {F.}~\bibnamefont {Dolde}},  \emph
  {et~al.},\ }\href@noop {} {\bibfield  {journal} {\bibinfo  {journal} {Nano
  Lett}\ }\textbf {\bibinfo {volume} {13}},\ \bibinfo {pages} {2738} (\bibinfo
  {year} {2013})}\BibitemShut {NoStop}%
\bibitem [{\citenamefont {Staudacher}\ \emph {et~al.}(2013)\citenamefont
  {Staudacher}, \citenamefont {Shi}, \citenamefont {Pezzagna} \emph
  {et~al.}}]{Joerg_nano_nmr}%
  \BibitemOpen
  \bibfield  {author} {\bibinfo {author} {\bibfnamefont {T.}~\bibnamefont
  {Staudacher}}, \bibinfo {author} {\bibfnamefont {F.}~\bibnamefont {Shi}},
  \bibinfo {author} {\bibfnamefont {S.}~\bibnamefont {Pezzagna}},  \emph
  {et~al.},\ }\href@noop {} {\bibfield  {journal} {\bibinfo  {journal}
  {Science}\ }\textbf {\bibinfo {volume} {339}},\ \bibinfo {pages} {561}
  (\bibinfo {year} {2013})}\BibitemShut {NoStop}%
\bibitem [{\citenamefont {Shi}\ \emph {et~al.}(2015)\citenamefont {Shi},
  \citenamefont {Zhang}, \citenamefont {Wang} \emph {et~al.}}]{single_esr}%
  \BibitemOpen
  \bibfield  {author} {\bibinfo {author} {\bibfnamefont {F.~Z.}\ \bibnamefont
  {Shi}}, \bibinfo {author} {\bibfnamefont {Q.}~\bibnamefont {Zhang}}, \bibinfo
  {author} {\bibfnamefont {P.~F.}\ \bibnamefont {Wang}},  \emph {et~al.},\
  }\href@noop {} {\bibfield  {journal} {\bibinfo  {journal} {Science}\ }\textbf
  {\bibinfo {volume} {347}},\ \bibinfo {pages} {1135} (\bibinfo {year}
  {2015})}\BibitemShut {NoStop}%
\bibitem [{\citenamefont {Tetienne}\ \emph
  {et~al.}(2014{\natexlab{a}})\citenamefont {Tetienne}, \citenamefont
  {Hingant}, \citenamefont {Kim} \emph {et~al.}}]{quimag}%
  \BibitemOpen
  \bibfield  {author} {\bibinfo {author} {\bibfnamefont {J.~P.}\ \bibnamefont
  {Tetienne}}, \bibinfo {author} {\bibfnamefont {T.}~\bibnamefont {Hingant}},
  \bibinfo {author} {\bibfnamefont {J.~V.}\ \bibnamefont {Kim}},  \emph
  {et~al.},\ }\href@noop {} {\bibfield  {journal} {\bibinfo  {journal}
  {Science}\ }\textbf {\bibinfo {volume} {344}},\ \bibinfo {pages} {1366}
  (\bibinfo {year} {2014}{\natexlab{a}})}\BibitemShut {NoStop}%
\bibitem [{\citenamefont {Degen}\ \emph {et~al.}(2009)\citenamefont {Degen},
  \citenamefont {Poggio}, \citenamefont {Mamin} \emph {et~al.}}]{nmr_imaging}%
  \BibitemOpen
  \bibfield  {author} {\bibinfo {author} {\bibfnamefont {C.~L.}\ \bibnamefont
  {Degen}}, \bibinfo {author} {\bibfnamefont {M.}~\bibnamefont {Poggio}},
  \bibinfo {author} {\bibfnamefont {H.~J.}\ \bibnamefont {Mamin}},  \emph
  {et~al.},\ }\href@noop {} {\bibfield  {journal} {\bibinfo  {journal} {P Natl
  Acad Sci USA}\ }\textbf {\bibinfo {volume} {106}},\ \bibinfo {pages} {1313}
  (\bibinfo {year} {2009})}\BibitemShut {NoStop}%
\bibitem [{\citenamefont {Mamin}\ \emph {et~al.}(2007)\citenamefont {Mamin},
  \citenamefont {Poggio}, \citenamefont {Degen},\ and\ \citenamefont
  {Rugar}}]{nmr_imaging_rugar2007}%
  \BibitemOpen
  \bibfield  {author} {\bibinfo {author} {\bibfnamefont {H.~J.}\ \bibnamefont
  {Mamin}}, \bibinfo {author} {\bibfnamefont {M.}~\bibnamefont {Poggio}},
  \bibinfo {author} {\bibfnamefont {C.~L.}\ \bibnamefont {Degen}}, \ and\
  \bibinfo {author} {\bibfnamefont {D.}~\bibnamefont {Rugar}},\ }\href@noop {}
  {\bibfield  {journal} {\bibinfo  {journal} {Nat Nanotechnol}\ }\textbf
  {\bibinfo {volume} {2}},\ \bibinfo {pages} {301} (\bibinfo {year}
  {2007})}\BibitemShut {NoStop}%
\bibitem [{\citenamefont {Giovannetti}\ \emph {et~al.}(2011)\citenamefont
  {Giovannetti}, \citenamefont {Lloyd},\ and\ \citenamefont
  {Maccone}}]{qmetro_review}%
  \BibitemOpen
  \bibfield  {author} {\bibinfo {author} {\bibfnamefont {V.}~\bibnamefont
  {Giovannetti}}, \bibinfo {author} {\bibfnamefont {S.}~\bibnamefont {Lloyd}},
  \ and\ \bibinfo {author} {\bibfnamefont {L.}~\bibnamefont {Maccone}},\
  }\href@noop {} {\bibfield  {journal} {\bibinfo  {journal} {Nat Photonics}\
  }\textbf {\bibinfo {volume} {5}},\ \bibinfo {pages} {222} (\bibinfo {year}
  {2011})}\BibitemShut {NoStop}%
\bibitem [{\citenamefont {Du}\ \emph {et~al.}(2009)\citenamefont {Du},
  \citenamefont {Rong}, \citenamefont {Zhao} \emph {et~al.}}]{du_nature_2009}%
  \BibitemOpen
  \bibfield  {author} {\bibinfo {author} {\bibfnamefont {J.~F.}\ \bibnamefont
  {Du}}, \bibinfo {author} {\bibfnamefont {X.}~\bibnamefont {Rong}}, \bibinfo
  {author} {\bibfnamefont {N.}~\bibnamefont {Zhao}},  \emph {et~al.},\
  }\href@noop {} {\bibfield  {journal} {\bibinfo  {journal} {Nature}\ }\textbf
  {\bibinfo {volume} {461}},\ \bibinfo {pages} {1265} (\bibinfo {year}
  {2009})}\BibitemShut {NoStop}%
\bibitem [{\citenamefont {Ryan}\ \emph {et~al.}(2010)\citenamefont {Ryan},
  \citenamefont {Hodges},\ and\ \citenamefont {Cory}}]{dd_nv}%
  \BibitemOpen
  \bibfield  {author} {\bibinfo {author} {\bibfnamefont {C.~A.}\ \bibnamefont
  {Ryan}}, \bibinfo {author} {\bibfnamefont {J.~S.}\ \bibnamefont {Hodges}}, \
  and\ \bibinfo {author} {\bibfnamefont {D.~G.}\ \bibnamefont {Cory}},\
  }\href@noop {} {\bibfield  {journal} {\bibinfo  {journal} {Phys Rev Lett}\
  }\textbf {\bibinfo {volume} {105}} (\bibinfo {year} {2010})}\BibitemShut
  {NoStop}%
\bibitem [{\citenamefont {Hall}\ \emph
  {et~al.}(2010{\natexlab{a}})\citenamefont {Hall}, \citenamefont {Hill},
  \citenamefont {Cole},\ and\ \citenamefont {Hollenberg}}]{opt_dd_nv}%
  \BibitemOpen
  \bibfield  {author} {\bibinfo {author} {\bibfnamefont {L.~T.}\ \bibnamefont
  {Hall}}, \bibinfo {author} {\bibfnamefont {C.~D.}\ \bibnamefont {Hill}},
  \bibinfo {author} {\bibfnamefont {J.~H.}\ \bibnamefont {Cole}}, \ and\
  \bibinfo {author} {\bibfnamefont {L.~C.}\ \bibnamefont {Hollenberg}},\
  }\href@noop {} {\bibfield  {journal} {\bibinfo  {journal} {Phys Rev B}\
  }\textbf {\bibinfo {volume} {82}},\ \bibinfo {pages} {045208} (\bibinfo
  {year} {2010}{\natexlab{a}})}\BibitemShut {NoStop}%
\bibitem [{\citenamefont {Naydenov}\ \emph {et~al.}(2011)\citenamefont
  {Naydenov}, \citenamefont {Dolde}, \citenamefont {Hall} \emph
  {et~al.}}]{dd_nv_Joerg}%
  \BibitemOpen
  \bibfield  {author} {\bibinfo {author} {\bibfnamefont {B.}~\bibnamefont
  {Naydenov}}, \bibinfo {author} {\bibfnamefont {F.}~\bibnamefont {Dolde}},
  \bibinfo {author} {\bibfnamefont {L.~T.}\ \bibnamefont {Hall}},  \emph
  {et~al.},\ }\href@noop {} {\bibfield  {journal} {\bibinfo  {journal} {Phys
  Rev B}\ }\textbf {\bibinfo {volume} {83}},\ \bibinfo {pages} {081201}
  (\bibinfo {year} {2011})}\BibitemShut {NoStop}%
\bibitem [{\citenamefont {De~Lange}\ \emph {et~al.}(2011)\citenamefont
  {De~Lange}, \citenamefont {Riste}, \citenamefont {Dobrovitski},\ and\
  \citenamefont {Hanson}}]{dd_nv_hanson}%
  \BibitemOpen
  \bibfield  {author} {\bibinfo {author} {\bibfnamefont {G.}~\bibnamefont
  {De~Lange}}, \bibinfo {author} {\bibfnamefont {D.}~\bibnamefont {Riste}},
  \bibinfo {author} {\bibfnamefont {V.}~\bibnamefont {Dobrovitski}}, \ and\
  \bibinfo {author} {\bibfnamefont {R.}~\bibnamefont {Hanson}},\ }\href@noop {}
  {\bibfield  {journal} {\bibinfo  {journal} {Phys Rev Lett}\ }\textbf
  {\bibinfo {volume} {106}},\ \bibinfo {pages} {080802} (\bibinfo {year}
  {2011})}\BibitemShut {NoStop}%
\bibitem [{\citenamefont {Bylander}\ \emph {et~al.}(2011)\citenamefont
  {Bylander}, \citenamefont {Gustavsson}, \citenamefont {Yan} \emph
  {et~al.}}]{noise_spectrum}%
  \BibitemOpen
  \bibfield  {author} {\bibinfo {author} {\bibfnamefont {J.}~\bibnamefont
  {Bylander}}, \bibinfo {author} {\bibfnamefont {S.}~\bibnamefont
  {Gustavsson}}, \bibinfo {author} {\bibfnamefont {F.}~\bibnamefont {Yan}},
  \emph {et~al.},\ }\href@noop {} {\bibfield  {journal} {\bibinfo  {journal}
  {Nat. Phys.}\ }\textbf {\bibinfo {volume} {7}},\ \bibinfo {pages} {565}
  (\bibinfo {year} {2011})}\BibitemShut {NoStop}%
\bibitem [{\citenamefont {?lvarez}\ and\ \citenamefont
  {Suter}(2011)}]{noise_spectrum_suter}%
  \BibitemOpen
  \bibfield  {author} {\bibinfo {author} {\bibfnamefont {G.~A.}\ \bibnamefont
  {\'Alvarez}}\ and\ \bibinfo {author} {\bibfnamefont {D.}~\bibnamefont
  {Suter}},\ }\href@noop {} {\bibfield  {journal} {\bibinfo  {journal} {Phys
  Rev Lett}\ }\textbf {\bibinfo {volume} {107}},\ \bibinfo {pages} {230501}
  (\bibinfo {year} {2011})}\BibitemShut {NoStop}%
\bibitem [{\citenamefont {Kotler}\ \emph {et~al.}(2011)\citenamefont {Kotler},
  \citenamefont {Akerman}, \citenamefont {Glickman} \emph {et~al.}}]{qu_lia}%
  \BibitemOpen
  \bibfield  {author} {\bibinfo {author} {\bibfnamefont {S.}~\bibnamefont
  {Kotler}}, \bibinfo {author} {\bibfnamefont {N.}~\bibnamefont {Akerman}},
  \bibinfo {author} {\bibfnamefont {Y.}~\bibnamefont {Glickman}},  \emph
  {et~al.},\ }\href@noop {} {\bibfield  {journal} {\bibinfo  {journal}
  {Nature}\ }\textbf {\bibinfo {volume} {473}},\ \bibinfo {pages} {61}
  (\bibinfo {year} {2011})}\BibitemShut {NoStop}%
\bibitem [{\citenamefont {Cooper}\ \emph {et~al.}(2014)\citenamefont {Cooper},
  \citenamefont {Magesan}, \citenamefont {Yum},\ and\ \citenamefont
  {Cappellaro}}]{walsh_sensing}%
  \BibitemOpen
  \bibfield  {author} {\bibinfo {author} {\bibfnamefont {A.}~\bibnamefont
  {Cooper}}, \bibinfo {author} {\bibfnamefont {E.}~\bibnamefont {Magesan}},
  \bibinfo {author} {\bibfnamefont {H.~N.}\ \bibnamefont {Yum}}, \ and\
  \bibinfo {author} {\bibfnamefont {P.}~\bibnamefont {Cappellaro}},\
  }\href@noop {} {\bibfield  {journal} {\bibinfo  {journal} {Nature
  Communications}\ }\textbf {\bibinfo {volume} {5}},\ \bibinfo {pages} {3141}
  (\bibinfo {year} {2014})}\BibitemShut {NoStop}%
\bibitem [{\citenamefont {Shore}(1973)}]{haar_appl}%
  \BibitemOpen
  \bibfield  {author} {\bibinfo {author} {\bibfnamefont {J.~E.}\ \bibnamefont
  {Shore}},\ }\href@noop {} {\bibfield  {journal} {\bibinfo  {journal}
  {Communications, IEEE Transactions on}\ }\textbf {\bibinfo {volume} {21}},\
  \bibinfo {pages} {209} (\bibinfo {year} {1973})}\BibitemShut {NoStop}%
\bibitem [{\citenamefont {Stankovi?}\ and\ \citenamefont
  {Falkowski}(2003)}]{haar_status}%
  \BibitemOpen
  \bibfield  {author} {\bibinfo {author} {\bibfnamefont {R.~S.}\ \bibnamefont
  {Stankovi?}}\ and\ \bibinfo {author} {\bibfnamefont {B.~J.}\ \bibnamefont
  {Falkowski}},\ }\href@noop {} {\bibfield  {journal} {\bibinfo  {journal}
  {Computers and Electrical Engineering}\ }\textbf {\bibinfo {volume} {29}},\
  \bibinfo {pages} {25} (\bibinfo {year} {2003})}\BibitemShut {NoStop}%
\bibitem [{\citenamefont {Daubechies}(1992)}]{wavelet_textbook}%
  \BibitemOpen
  \bibfield  {author} {\bibinfo {author} {\bibfnamefont {I.}~\bibnamefont
  {Daubechies}},\ }\href@noop {} {\emph {\bibinfo {title} {Ten lectures on
  wavelets}}},\ Vol.~\bibinfo {volume} {61}\ (\bibinfo  {publisher} {SIAM},\
  \bibinfo {year} {1992})\BibitemShut {NoStop}%
\bibitem [{\citenamefont {Hunter}(2007)}]{matplotlib}%
  \BibitemOpen
  \bibfield  {author} {\bibinfo {author} {\bibfnamefont {J.~D.}\ \bibnamefont
  {Hunter}},\ }\href@noop {} {\bibfield  {journal} {\bibinfo  {journal}
  {Computing in science and engineering}\ }\textbf {\bibinfo {volume} {9}},\
  \bibinfo {pages} {90} (\bibinfo {year} {2007})}\BibitemShut {NoStop}%
\bibitem [{\citenamefont {Hines}\ and\ \citenamefont
  {Carnevale}(1997)}]{neuron_sim}%
  \BibitemOpen
  \bibfield  {author} {\bibinfo {author} {\bibfnamefont {M.~L.}\ \bibnamefont
  {Hines}}\ and\ \bibinfo {author} {\bibfnamefont {N.~T.}\ \bibnamefont
  {Carnevale}},\ }\href@noop {} {\bibfield  {journal} {\bibinfo  {journal}
  {Neural computation}\ }\textbf {\bibinfo {volume} {9}},\ \bibinfo {pages}
  {1179} (\bibinfo {year} {1997})}\BibitemShut {NoStop}%
\bibitem [{\citenamefont {Degen}(2008)}]{qimag_degen}%
  \BibitemOpen
  \bibfield  {author} {\bibinfo {author} {\bibfnamefont {C.~L.}\ \bibnamefont
  {Degen}},\ }\href@noop {} {\bibfield  {journal} {\bibinfo  {journal} {Appl
  Phys Lett}\ }\textbf {\bibinfo {volume} {92}},\ \bibinfo {pages} {243111}
  (\bibinfo {year} {2008})}\BibitemShut {NoStop}%
\bibitem [{\citenamefont {Tetienne}\ \emph
  {et~al.}(2014{\natexlab{b}})\citenamefont {Tetienne}, \citenamefont
  {Hingant}, \citenamefont {Kim} \emph {et~al.}}]{qimag_domainwall}%
  \BibitemOpen
  \bibfield  {author} {\bibinfo {author} {\bibfnamefont {J.-P.}\ \bibnamefont
  {Tetienne}}, \bibinfo {author} {\bibfnamefont {T.}~\bibnamefont {Hingant}},
  \bibinfo {author} {\bibfnamefont {J.-V.}\ \bibnamefont {Kim}},  \emph
  {et~al.},\ }\href@noop {} {\bibfield  {journal} {\bibinfo  {journal}
  {Science}\ }\textbf {\bibinfo {volume} {344}},\ \bibinfo {pages} {1366}
  (\bibinfo {year} {2014}{\natexlab{b}})}\BibitemShut {NoStop}%
\bibitem [{\citenamefont {Maletinsky}\ \emph {et~al.}(2012)\citenamefont
  {Maletinsky}, \citenamefont {Hong}, \citenamefont {Grinolds} \emph
  {et~al.}}]{lukin_imaging}%
  \BibitemOpen
  \bibfield  {author} {\bibinfo {author} {\bibfnamefont {P.}~\bibnamefont
  {Maletinsky}}, \bibinfo {author} {\bibfnamefont {S.}~\bibnamefont {Hong}},
  \bibinfo {author} {\bibfnamefont {M.~S.}\ \bibnamefont {Grinolds}},  \emph
  {et~al.},\ }\href@noop {} {\bibfield  {journal} {\bibinfo  {journal} {Nat
  Nanotechnol}\ }\textbf {\bibinfo {volume} {7}},\ \bibinfo {pages} {320}
  (\bibinfo {year} {2012})}\BibitemShut {NoStop}%
\bibitem [{\citenamefont {Grinolds}\ \emph {et~al.}(2013)\citenamefont
  {Grinolds}, \citenamefont {Hong}, \citenamefont {Maletinsky} \emph
  {et~al.}}]{yacoby_imaging}%
  \BibitemOpen
  \bibfield  {author} {\bibinfo {author} {\bibfnamefont {M.~S.}\ \bibnamefont
  {Grinolds}}, \bibinfo {author} {\bibfnamefont {S.}~\bibnamefont {Hong}},
  \bibinfo {author} {\bibfnamefont {P.}~\bibnamefont {Maletinsky}},  \emph
  {et~al.},\ }\href@noop {} {\bibfield  {journal} {\bibinfo  {journal} {Nat.
  Phys.}\ }\textbf {\bibinfo {volume} {9}},\ \bibinfo {pages} {215} (\bibinfo
  {year} {2013})}\BibitemShut {NoStop}%
\bibitem [{\citenamefont {Maurer}\ \emph {et~al.}(2010)\citenamefont {Maurer},
  \citenamefont {Maze}, \citenamefont {Stanwix} \emph
  {et~al.}}]{lukin_farfield-imaging}%
  \BibitemOpen
  \bibfield  {author} {\bibinfo {author} {\bibfnamefont {P.~C.}\ \bibnamefont
  {Maurer}}, \bibinfo {author} {\bibfnamefont {J.~R.}\ \bibnamefont {Maze}},
  \bibinfo {author} {\bibfnamefont {P.~L.}\ \bibnamefont {Stanwix}},  \emph
  {et~al.},\ }\href@noop {} {\bibfield  {journal} {\bibinfo  {journal} {Nat.
  Phys.}\ }\textbf {\bibinfo {volume} {6}},\ \bibinfo {pages} {912} (\bibinfo
  {year} {2010})}\BibitemShut {NoStop}%
\bibitem [{\citenamefont {Steinert}\ \emph {et~al.}(2013)\citenamefont
  {Steinert}, \citenamefont {Ziem}, \citenamefont {Hall} \emph
  {et~al.}}]{joerg_imaging2013}%
  \BibitemOpen
  \bibfield  {author} {\bibinfo {author} {\bibfnamefont {S.}~\bibnamefont
  {Steinert}}, \bibinfo {author} {\bibfnamefont {F.}~\bibnamefont {Ziem}},
  \bibinfo {author} {\bibfnamefont {L.~T.}\ \bibnamefont {Hall}},  \emph
  {et~al.},\ }\href@noop {} {\bibfield  {journal} {\bibinfo  {journal} {Nature
  Communications}\ }\textbf {\bibinfo {volume} {4}} (\bibinfo {year}
  {2013})}\BibitemShut {NoStop}%
\bibitem [{\citenamefont {Haberle}\ \emph {et~al.}(2013)\citenamefont
  {Haberle}, \citenamefont {Schmid-Lorch}, \citenamefont {Karrai} \emph
  {et~al.}}]{high_dynrange_imaging}%
  \BibitemOpen
  \bibfield  {author} {\bibinfo {author} {\bibfnamefont {T.}~\bibnamefont
  {Haberle}}, \bibinfo {author} {\bibfnamefont {D.}~\bibnamefont
  {Schmid-Lorch}}, \bibinfo {author} {\bibfnamefont {K.}~\bibnamefont
  {Karrai}},  \emph {et~al.},\ }\href@noop {} {\bibfield  {journal} {\bibinfo
  {journal} {Phys Rev Lett}\ }\textbf {\bibinfo {volume} {111}} (\bibinfo
  {year} {2013})}\BibitemShut {NoStop}%
\bibitem [{\citenamefont {Lazariev}\ and\ \citenamefont
  {Balasubramanian}(2015)}]{nv_imag_projreconstr}%
  \BibitemOpen
  \bibfield  {author} {\bibinfo {author} {\bibfnamefont {A.}~\bibnamefont
  {Lazariev}}\ and\ \bibinfo {author} {\bibfnamefont {G.}~\bibnamefont
  {Balasubramanian}},\ }\href@noop {} {\bibfield  {journal} {\bibinfo
  {journal} {arXiv preprint arXiv:1505.02904}\ } (\bibinfo {year}
  {2015})}\BibitemShut {NoStop}%
\bibitem [{\citenamefont {Hall}\ \emph {et~al.}(2009)\citenamefont {Hall},
  \citenamefont {Cole}, \citenamefont {Hill},\ and\ \citenamefont
  {Hollenberg}}]{nv_fluctmag}%
  \BibitemOpen
  \bibfield  {author} {\bibinfo {author} {\bibfnamefont {L.~T.}\ \bibnamefont
  {Hall}}, \bibinfo {author} {\bibfnamefont {J.~H.}\ \bibnamefont {Cole}},
  \bibinfo {author} {\bibfnamefont {C.~D.}\ \bibnamefont {Hill}}, \ and\
  \bibinfo {author} {\bibfnamefont {L.~C.~L.}\ \bibnamefont {Hollenberg}},\
  }\href@noop {} {\bibfield  {journal} {\bibinfo  {journal} {Phys Rev Lett}\
  }\textbf {\bibinfo {volume} {103}},\ \bibinfo {pages} {220802} (\bibinfo
  {year} {2009})}\BibitemShut {NoStop}%
\bibitem [{\citenamefont {Hall}\ \emph
  {et~al.}(2010{\natexlab{b}})\citenamefont {Hall}, \citenamefont {Hill},
  \citenamefont {Cole} \emph {et~al.}}]{nv_ionch}%
  \BibitemOpen
  \bibfield  {author} {\bibinfo {author} {\bibfnamefont {L.~T.}\ \bibnamefont
  {Hall}}, \bibinfo {author} {\bibfnamefont {C.~D.}\ \bibnamefont {Hill}},
  \bibinfo {author} {\bibfnamefont {J.~H.}\ \bibnamefont {Cole}},  \emph
  {et~al.},\ }\href@noop {} {\bibfield  {journal} {\bibinfo  {journal}
  {Proceedings of the National Academy of Sciences}\ }\textbf {\bibinfo
  {volume} {107}},\ \bibinfo {pages} {18777} (\bibinfo {year}
  {2010}{\natexlab{b}})}\BibitemShut {NoStop}%
\bibitem [{\citenamefont {Hall}\ \emph {et~al.}(2012)\citenamefont {Hall},
  \citenamefont {Beart}, \citenamefont {Thomas} \emph {et~al.}}]{nv_neuron}%
  \BibitemOpen
  \bibfield  {author} {\bibinfo {author} {\bibfnamefont {L.~T.}\ \bibnamefont
  {Hall}}, \bibinfo {author} {\bibfnamefont {G.~C.~G.}\ \bibnamefont {Beart}},
  \bibinfo {author} {\bibfnamefont {E.~A.}\ \bibnamefont {Thomas}},  \emph
  {et~al.},\ }\href@noop {} {\bibfield  {journal} {\bibinfo  {journal}
  {Scientific Reports}\ }\textbf {\bibinfo {volume} {2}},\ \bibinfo {pages}
  {401} (\bibinfo {year} {2012})}\BibitemShut {NoStop}%
\bibitem [{\citenamefont {Pham}\ \emph {et~al.}(2011)\citenamefont {Pham},
  \citenamefont {Le~Sage}, \citenamefont {Stanwix} \emph
  {et~al.}}]{nv_imag_ensem}%
  \BibitemOpen
  \bibfield  {author} {\bibinfo {author} {\bibfnamefont {L.~M.}\ \bibnamefont
  {Pham}}, \bibinfo {author} {\bibfnamefont {D.}~\bibnamefont {Le~Sage}},
  \bibinfo {author} {\bibfnamefont {P.~L.}\ \bibnamefont {Stanwix}},  \emph
  {et~al.},\ }\href@noop {} {\bibfield  {journal} {\bibinfo  {journal} {New J
  Phys}\ }\textbf {\bibinfo {volume} {13}},\ \bibinfo {pages} {045021}
  (\bibinfo {year} {2011})}\BibitemShut {NoStop}%
\bibitem [{\citenamefont {Wolf}\ \emph {et~al.}(2015)\citenamefont {Wolf},
  \citenamefont {Neumann}, \citenamefont {Nakamura} \emph
  {et~al.}}]{nv_ensemble_magnetometer}%
  \BibitemOpen
  \bibfield  {author} {\bibinfo {author} {\bibfnamefont {T.}~\bibnamefont
  {Wolf}}, \bibinfo {author} {\bibfnamefont {P.}~\bibnamefont {Neumann}},
  \bibinfo {author} {\bibfnamefont {K.}~\bibnamefont {Nakamura}},  \emph
  {et~al.},\ }\href@noop {} {\bibfield  {journal} {\bibinfo  {journal} {Phys
  Rev X}\ }\textbf {\bibinfo {volume} {5}},\ \bibinfo {pages} {041001}
  (\bibinfo {year} {2015})}\BibitemShut {NoStop}%
\bibitem [{\citenamefont {Jensen}\ \emph {et~al.}(2014)\citenamefont {Jensen},
  \citenamefont {Leefer}, \citenamefont {Jarmola} \emph
  {et~al.}}]{budker_cavity}%
  \BibitemOpen
  \bibfield  {author} {\bibinfo {author} {\bibfnamefont {K.}~\bibnamefont
  {Jensen}}, \bibinfo {author} {\bibfnamefont {N.}~\bibnamefont {Leefer}},
  \bibinfo {author} {\bibfnamefont {A.}~\bibnamefont {Jarmola}},  \emph
  {et~al.},\ }\href@noop {} {\bibfield  {journal} {\bibinfo  {journal} {Phys
  Rev Lett}\ }\textbf {\bibinfo {volume} {112}} (\bibinfo {year}
  {2014})}\BibitemShut {NoStop}%
\end{thebibliography}
%\bibliographystyle{apsrev4-1}
%

\end{document}